# Supervised Speech Representation Learning for Parkinson's Disease Classification

*Parvaneh Janbakhshi[1,2], Ina Kodrasi[1]*

[1]Idiap Research Institute, Martigny, Switzerland
[2]École Polytechnique Fédérale de Lausanne, Lausanne, Switzerland
Email: {parvaneh.janbakhshi,ina.kodrasi}@idiap.ch

## Abstract

Recently proposed automatic pathological speech classification techniques use unsupervised auto-encoders to obtain a high-level abstract representation of speech. Since these representations are learned based on reconstructing the input, there is no guarantee that they are robust to pathology-unrelated cues such as speaker identity information. Further, these representations are not necessarily discriminative for pathology detection. In this paper, we exploit supervised auto-encoders to extract robust and discriminative speech representations for Parkinson's disease classification. To reduce the influence of speaker variabilities unrelated to pathology, we propose to obtain speaker identity-invariant representations by adversarial training of an auto-encoder and a speaker identification task. To obtain a discriminative representation, we propose to jointly train an auto-encoder and a pathological speech classifier. Experimental results on a Spanish database show that the proposed supervised representation learning methods yield more robust and discriminative representations for automatically classifying Parkinson's disease speech, outperforming the baseline unsupervised representation learning system.

## 1 Introduction

Parkinson's disease (PD) is a neurodegenerative disorder that disrupts the speech production mechanism resulting in hypokinetic dysarthria of speech. Hypokinetic dysarthria is characterized by imprecise articulation, abnormal speech rhythm, prosodic insufficiency, reduced stress, monoloudness, and breathiness [1, 2]. For diagnosis, management, and treatment of these speech deficits associated with PD, speech screening through clinical auditory-perceptual assessments is typically used. Such clinical assessments can be time-consuming, expensive, and inconsistent, since they are subjective and influenced by the level of expertise of clinicians.

To assist clinical speech screenings, a wide range of automatic PD speech classification techniques have been proposed [3–9]. The majority of state-of-the-art contributions are based on classical machine learning approaches, i.e., they extract handcrafted acoustic features and train classical classifiers on these handcrafted features to achieve pathological and neurotypical speech discrimination [5, 6]. Typically used acoustic features are inspired by clinicians' knowledge and aim to characterize different impaired speech dimensions, with e.g. Mel frequency cepstral coefficients aiming to characterize imprecise articulation, spectro-temporal sparsity features aiming to characterize breathiness, or rhythm-based features aiming to characterize abnormal rhythmic patterns [8–15]. Although handcrafted acoustic features have shown promising results, such features may fail to adequately capture pathological speech characteristics. Further, since handcrafted features are based on clinicians' knowledge, they may also fail to characterize abstract but important acoustic cues present in pathological speech.

As an alternative to using handcrafted acoustic features, high-level representations of speech can be extracted using data-driven deep learning approaches [7, 16–19]. The main challenge in successfully learning such representations is being able to systematically guide networks to learn robust and relevant features for pathological speech detection, while using the small amount of pathological training data that is typically available. To this end, long short-term memory Siamese networks trained on pairs of input data with the same phonetic content are used for dysarthric speech detection in [17]. Pairwise training guides the network to extract features that are discriminative of dysarthria while being robust to other unrelated speaker variabilities. However, since input data needs to have the same phonetic content, different networks need to be trained for different utterances. Exploiting pairwise training while using a single network for different utterances, a pairwise distance-based architecture has been proposed in [7]. Although promising results have been achieved in [7, 17], such architectures rely on having access to utterances with the same phonetic content from both neurotypical and dysarthric speakers.

Recently it has been proposed to learn high-level (but not necessarily robust and discriminative as explained in the following) representations through unsupervised auto-encoders operating on phonetically unmatched speech segments [18, 19]. In [18], representations are first extracted using auto-encoders trained on a large amount of neurotypical speech, while stacked auto-encoders are exploited in [19]. The extracted representations are then used as input for training PD classifiers. Unsupervised representation learning based on auto-encoders yields representations that are designed to reconstruct the input. Consequently, there is no guarantee that these learned representations are robust to pathology-unrelated cues such as acoustic information about the speaker identity. In addition, there is no guarantee that these representations are discriminative for pathology detection. To tackle these issues, in this paper we propose two methods to extract robust and discriminative representations from speech spectrograms exploiting supervised auto-encoders.

First, we propose to supervise the representation learning process such that only speaker-invariant information is retained. This is achieved through training an adversarial network by jointly minimizing the auto-encoder reconstruction loss and the performance of a (neurotypical) speaker identification (ID) task. The prominence of speaker variabilities unrelated to PD in such representations will be limited, and hence, it can be expected that the performance of PD classification can be improved. Suppressing unrelated speaker variabilities from representations in an adversarial training framework has been recently shown to improve the performance for different classification tasks such as speech emotion classification, phoneme/senone discrimination, and speaker de-identification [20–23].

Second, to ensure that the learned representations retain PD discriminative information, we propose to train the representation layer by jointly minimizing the auto-encoder reconstruction loss and maximizing the performance of PD classification. In [24] it has been shown that such supervised auto-encoders typically do not harm the performance compared to a standard neural network, since the incorporation of the reconstruction loss into the training procedure acts as a regularisation method. It should be noted that such a joint training procedure to learn discriminative representations for dysarthric speech classification has been investigated in [25], where however two encoders are used, i.e., an audio and a text encoder. Differently from [25] and inline with [18, 19], a single encoder is used in this paper.

Experimental results on a Spanish database of neurotypical and PD speakers show that using speaker-invariant and/or PD discriminative representations improves the PD classification performance compared to using representations learned in an unsupervised manner.

## 2 Technical Approach

Figure 1 illustrates the proposed representation learning for PD classification using an auto-encoder and two auxiliary modules, i.e., an adversarial speaker ID module and a PD classifier module. To obtain a speaker identity-invariant representation, the auto-encoder can be jointly trained with the speaker ID task in an adversarial manner (cf. Section 2.2). To obtain a PD discriminative representation, the auto-encoder can be jointly trained with the PD classifier (cf. Section 2.3). To obtain a speaker identity-invariant and PD discriminative representation, the auto-encoder can be jointly trained with both auxiliary tasks (cf. Section 2.4).

### 2.1 Auto-encoder

Similarly to [18], we consider a Convolutional Neural Network (CNN)-based auto-encoder to compute low-dimensional representations from chunks of speech spectrograms. Spectrograms are encoded with four convolutional layers (filter size: $3 \times 3$, stride: 1), with the number of feature maps on each layer being twice the number of feature maps on the previous layer (starting with 16 maps in the first layer). Each convolutional layer is followed by max-pooling (filter size: $2 \times 2$, stride: 2), batch normalization, and leaky ReLU activation functions. The output of the last convolutional layer is further processed with a fully connected layer (with 256 hidden units) to form the final feature representation, i.e., bottleneck representation, of size 128. The bottleneck representation is decoded into a reconstructed version of the input spectrograms by the decoder. The decoder components are stacked in reverse order of the encoder components, where transposed convolutional and interpolation layers are used instead of convolutional and max-pooling layers. In the remainder of this paper, the parameters of the encoder and decoder are denoted by $\theta_e$ and $\theta_d$ respectively.

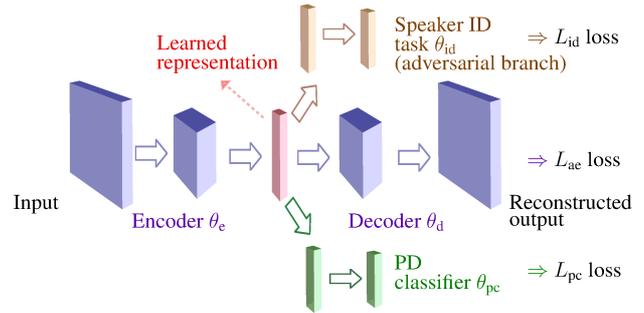

Figure 1: Proposed supervised representation learning for PD classification using an auto-encoder and auxiliary tasks. The auto-encoder is jointly trained with the auxiliary speaker ID task and/or with the auxiliary PD classifier.

### 2.2 Speaker identity-invariant representation with adversarial training

To learn representations robust to speaker variabilities unrelated to PD, i.e., speaker identity, the bottleneck representation of the auto-encoder in Section 2.1 is connected to a speaker ID module. The architecture of this module is adapted from the final classifier used in [18] and consists of two fully connected layers with 64 hidden units each, a leaky ReLU activation function after the first layer, and a Softmax activation function after the final (i.e., second) layer. The number of output units, i.e., the number of units in the final layer, is the same as the number of speakers used for the speaker ID task (cf. Section 3.2). To avoid over-fitting, a dropout layer with a rate of 0.2 is included between the bottleneck layer and the speaker ID module. The parameters of this module are denoted by $\theta_{id}$.

To obtain a compact representation where the information related to the speaker identity is minimized, we use adversarial training by minimizing the auto-encoder reconstruction loss $L_{ae}$ such that a low reconstruction error is achieved, while maximizing the speaker ID loss $L_{id}$ such that a low speaker ID accuracy is achieved. Adversarial training is achieved through the min-max optimization objective

$$(\hat{\theta}_e, \hat{\theta}_d, \hat{\theta}_{id}) = \arg\min_{\theta_e, \theta_d} \arg\max_{\theta_{id}} E(\theta_e, \theta_d, \theta_{id}), \quad (1)$$

with

$$E(\theta_e, \theta_d, \theta_{id}) = (1 - \lambda)L_{ae}(\theta_e, \theta_d) - \lambda L_{id}(\theta_e, \theta_{id}), \quad (2)$$

where $0 < \lambda < 1$ is the trade-off parameter between the auto-encoder and the adversarial loss functions (cf. Section 3.2). In practice, the optimal parameters in (2) are approximated using an alternating training procedure, where in the first step, the auto-encoder parameters $\theta_e$ and $\theta_d$ are updated assuming fixed speaker ID parameters $\theta_{id}$, and in the second step, the parameters $\theta_{id}$ are updated assuming a fixed $\theta_e$ and $\theta_d$ obtained in the first step, i.e.,

$$(\hat{\theta}_e, \hat{\theta}_d) = \arg\min_{\theta_e, \theta_d} E(\theta_e, \theta_d, \hat{\theta}_{id}), \quad (3)$$

$$\hat{\theta}_{id} = \arg\max_{\theta_{id}} E(\hat{\theta}_e, \hat{\theta}_d, \theta_{id}). \quad (4)$$

Each parameter set is updated using Stochastic Gradient

Decent (SGD) as in [21]. While all training speakers (neurotypical and pathological) are used for optimizing the reconstruction loss $L_{ae}$, we consider data only from neurotypical speakers to optimize the speaker ID loss $L_{id}$. This ensures that only non-pathological speaker variabilities are suppressed from the bottleneck representation.

### 2.3 PD discriminative representation

To learn PD discriminative representations, the bottleneck representation of the auto-encoder in Section 2.1 is connected to a PD classifier module. The same architecture of fully connected layers as for the speaker ID module in Section 2.2 is used for the PD classifier module. However, differently from the speaker ID module, the final layer for the PD classifier module consists of 2 output units since we are dealing with binary classification (i.e., PD vs. neurotypical speech). The parameters of this module are denoted by $\theta_{pc}$.

The optimal parameters $\theta_e$, $\theta_d$, and $\theta_{pc}$ are computed as the ones simultaneously minimizing the auto-encoder reconstruction loss $L_{ae}$ and the PD classification loss $L_{pc}$, i.e.,

$$(\hat{\theta}_e, \hat{\theta}_d, \hat{\theta}_{pc}) = \arg\min_{\theta_e, \theta_d, \theta_{pc}} E(\theta_e, \theta_d, \theta_{pc}), \quad (5)$$

with

$$E(\theta_e, \theta_d, \theta_{pc}) = (1-\alpha)L_{ae}(\theta_e, \theta_d) + \alpha L_{pc}(\theta_e, \theta_{pc}), \quad (6)$$

where $0 < \alpha < 1$ is the trade-off parameter between the two loss functions (cf. Section 3.2). Similarly to before, the SGD algorithm is used for finding the optimal parameters.

### 2.4 Fusion

To jointly learn a speaker identity-invariant and PD discriminative representation, we also consider training the auto-encoder in Section 2.1 using both auxiliary modules in Sections 2.2 and 2.3 through the optimization objective

$$(\hat{\theta}_e, \hat{\theta}_d, \hat{\theta}_{pc}, \hat{\theta}_{id}) = \arg\min_{\theta_e, \theta_d, \theta_{pc}} \arg\max_{\theta_{id}} E(\theta_e, \theta_d, \theta_{pc}, \theta_{id}), \quad (7)$$

where

$$E(\theta_e, \theta_d, \theta_{pc}, \theta_{id}) = (1-\alpha-\lambda)L_{ae}(\theta_e, \theta_d) \\ + \alpha L_{pc}(\theta_e, \theta_{pc}) - \lambda L_{id}(\theta_e, \theta_{id}). \quad (8)$$

The solution to (7) is approximated using a similar alternating training procedure as in Section 2.2.

### 2.5 PD speech classification

After obtaining the bottleneck representation following any of the training procedures outlined in Sections 2.2, 2.3, or 2.4, this representation is used to train a PD speech classifier. The classifier architecture is identical to the auxiliary classifier module in Section 2.3. The final decision for an unseen (test) speaker is made by applying soft voting on the classifier prediction scores for all input spectrograms belonging to that speaker. The Pytorch [26] implementation of our approach is available online[1] for the research community.

---

[1] https://github.com/idiap/pddetection-reps-learning

## 3 Experimental Results

In this section, the performance of the PD speech classification system using the proposed supervised representation learning techniques is evaluated and compared to using the unsupervised learning baseline system from [18].

### 3.1 Database

We consider Spanish recordings from 50 PD patients (25 males, 25 females) and 50 neurotypical speakers (25 males, 25 females) from the PC-GITA database [27]. Each speaker utters 24 words, 10 sentences, and 1 text recorded at a sampling frequency of 44.1 kHz. After downsampling to 16 kHz, speech-only segments are manually extracted from the word recordings and using an energy-based voice activity detector for all other recordings [28]. The average length of the speech material considered for each speaker is 59.9 s.

### 3.2 Training, evaluation, and baseline system

As in [18], the input representations are Mel-scale representations of 500 ms segments of speech with 50% overlap. Mel-scale representations are computed using 32 ms Hamming windows with a frame shift of 4 ms and 126 Mel bands. Z-score normalization is applied to all input representations.

For training and evaluation, we use a stratified speaker-independent 10-fold cross-validation framework, i.e., there is no overlap of speakers across different folds. In each training fold, a development fold of the same size as the test fold is set aside for early-stopping. For the speaker ID auxiliary task, utterances from the neurotypical speakers of the training set (i.e., 45 speakers) are split without overlap into 60% train, 20% development, and 20% test sets. Cross-entropy is used for the auxiliary loss functions $L_{id}$ and $L_{pc}$, whereas mean square reconstruction error is used for the auto-encoder loss $L_{ae}$. The models are trained with a batch size of 128 and an initial learning rate of 0.02. The learning rate is halved each time the loss on the development set does not decrease for 5 consecutive iterations. Training is stopped either after 100 epochs or after the learning rate has decreased beyond 0.002.

To demonstrate the advantages of the obtained speaker identity-invariant and PD discriminative representations, we consider the system in [18] as the baseline system where the bottleneck representation is learned using an auto-encoder (with the same architecture as in Section 2.2) without any supervision. Furthermore, to investigate the suitability of supervised representation learning for suppressing irrelevant speaker identity information, we also train a speaker ID module on each of the learned representations. The architecture of this module is identical to the auxiliary speaker ID module in Section 2.2.

The PD classification performance is evaluated in terms of accuracy (i.e., percentage of correctly classified neurotypical and PD speakers) and the area under the ROC curve (AUC). The performance for the speaker ID task is evaluated for unseen (test) utterances also using accuracy (i.e., percentage of correctly identified speakers) and AUC. To reduce the impact of the random seed on the final model parameters, all networks are trained with 5 different random seeds. The reported performance measures are the mean and standard deviation of the performance obtained by models trained using different seeds.

Table 1: *Mean and standard deviation of the PD classification accuracy [%] and AUC score.*

| Auxiliary task in representation learning | Accuracy | AUC |
|---|---|---|
| No auxiliary task (baseline) | $66.20 \pm 1.17$ | $0.77 \pm 0.02$ |
| Adversarial speaker invariant training | $72.00 \pm 5.62$ | $\mathbf{0.84} \pm 0.04$ |
| PD discriminative training | $71.00 \pm 1.90$ | $0.78 \pm 0.02$ |
| Fusion (speaker invariant+PD discriminative training) | $\mathbf{75.4} \pm 1.02$ | $0.80 \pm 0.02$ |

Table 2: *Mean and standard deviation of the speaker ID classification accuracy [%] and AUC score.*

| Auxiliary task in representation learning | Accuracy | AUC |
|---|---|---|
| No auxiliary task (baseline) | $34.71 \pm 11.94$ | $0.90 \pm 0.06$ |
| Adversarial speaker invariant training | $2.31 \pm 0.27$ | $0.54 \pm 0.01$ |
| PD discriminative training | $18.15 \pm 14.27$ | $0.76 \pm 0.08$ |
| Fusion (speaker invariant+PD discriminative training) | $2.59 \pm 0.19$ | $0.58 \pm 0.02$ |

To select the hyper-parameters $\lambda$ and $\alpha$ of the proposed approach (cf. (2) and (6)), we use grid-search for the set of values $\lambda, \alpha \in \{0.01, 0.03, ..., 0.07\}$. The final hyper-parameters $\lambda$ and $\alpha$ are selected as the ones yielding the highest mean PD classification accuracy on the development set. The final selected value for both $\lambda$ and $\alpha$ using the grid-search is 0.01. It should be noted that hyper-parameters are optimized this way only when supervised learning is used with a single auxiliary task, i.e., the speaker ID task or the PD classifier. For the fusion approach in Section 2.4, the hyper-parameters used in (8) are not optimized but are set to the values obtained from their optimization on each of the individual tasks (i.e., $\lambda = \alpha = 0.01$).

### 3.3 Results

Table 1 presents the PD classification accuracy and AUC values obtained using the proposed supervised representations learned through auxiliary tasks and using the baseline representation from [18] learned without any supervision[2].

It can be observed that using the representations learned by any of the proposed auxiliary tasks improves the performance of PD classification compared to using the baseline unsupervised representation. When comparing the two proposed supervised representation learning approaches, a larger performance improvement is observed in terms of both performance measures for the speaker-invariant training. Furthermore, fusing both auxiliary tasks to obtain a robust and discriminative representation yields a better PD classification accuracy than other representations, clearly outperforming the unsupervised baseline system as well. It can be observed that while the fusion of auxiliary tasks improves the PD classification accuracy as opposed to using any of the auxiliary tasks, the resulting AUC is lower than when using adversarial speaker invariant training. We suspect this occurs due to the use of suboptimal hyper-parameters for the fusion of auxiliary tasks, while optimal hyper-parameters are used for the adversarial speaker invariant training.

In summary, the results presented in Table 1 confirm the advantages of supervised representation learning for PD classification. To investigate the suppression of irrelevant speaker identity information in each of the supervised representations as opposed to the unsupervised representation, Table 2 presents the accuracy and AUC values obtained for the speaker ID task on all the different representations. It can be observed that using the baseline (unsupervised) representation results in the highest speaker ID performance. This result confirms that unsupervised training yields representations containing speaker identity cues, reducing as a result the generalization and final performance of PD classification (cf. Table 1). Further, as expected, the lowest speaker ID performance is observed for the speaker identity-invariant representations obtained using adversarial training. These results confirm the suitability of adversarial training to reduce the presence of irrelevant speaker identity cues in the bottleneck representation. Finally, it can be observed that although the PD discriminative feature representation results in a higher speaker ID performance than adversarial training, it yields a significantly lower speaker ID performance than the unsupervised baseline representation. This result shows that supervising the auto-encoder training such that a discriminative feature representation for PD classification is learned, inherently reduces the presence of speaker identity cues, since they are irrelevant to the PD classification task.

## 4 Conclusion

In this paper, we proposed to use supervised representation learning frameworks with auxiliary tasks for PD classification. To obtain a representation that is robust to irrelevant speaker identity cues, we have trained an auto-encoder jointly with an auxiliary speaker ID task in an adversarial fashion. To obtain a representation that is discriminative for PD classification, we have trained an auto-encoder jointly with an auxiliary PD classifier. Experimental results on a Spanish database of neurotypical and PD speakers have shown that such speaker identity-invariant and PD discriminative representations are advantageous for PD classification, outperforming using representations learned in an unsupervised manner.

In the future, we plan to investigate the presence of other pathology-unrelated cues (e.g., age and gender) in the learned representations. We expect such cues to also be detrimental to PD classification performance, and hence, we plan to incorporate their suppression within the proposed adversarial training framework.

## Acknowledgment

The authors would like to acknowledge the support of the Swiss National Science Foundation project no CR-SII5_173711 "MoSpeeDi" on "*Motor Speech Disorders: characterizing phonetic speech planning and motor speech programming/execution and their impairments*".

---

[2] It should be noted that the auto-encoder used in [18] was trained on a larger neurotypical speech database. However, although not presented here due to space constraints, using the same neurotypical speech database for training the auto-encoder did not result in a better performance than the performance obtained using only the PC-GITA database.